\documentclass{article}

\usepackage{arxiv}

\usepackage[utf8]{inputenc} 
\usepackage[T1]{fontenc}    
\usepackage{hyperref}       
\usepackage{url}            
\usepackage{booktabs}       
\usepackage{amsfonts}       
\usepackage{nicefrac}       
\usepackage{microtype}      
\usepackage{graphicx}
\newcommand{\para}[1]{\vspace{0.5\baselineskip}\noindent\textbf{#1.~}}

\title{Characterizing IoT Data and its Quality for Use}

\author{
	Nashez Zubair\\
	Department of Computational and Data Sciences\\
	Indian Institute of Science\\
	Bangalore, India 560012 \\
	\texttt{nashezzubair@gmail.com} \\
	\And
	Niranjan A \\
	Department of Computational and Data Sciences\\
	Indian Institute of Science\\
	Bangalore, India 560012 \\
	\texttt{}\\
	\And
	Kiran Hebbar\\
	Department of Computational and Data Sciences\\
	Indian Institute of Science\\
	Bangalore, India 560012 \\
	\texttt{}\\
	\And
	Yogesh Simmhan \\
	Department of Computational and Data Sciences\\
	Indian Institute of Science\\
	Bangalore, India 560012 \\
	\texttt{simmhan@iisc.ac.in} \\
}

\begin{document}
\maketitle

\begin{abstract}
The Internet of Things (IoT) is a cyber physical social system that encompasses science, enterprise and societal domains. Data is the most important commodity in IoT, enabling the ``smarts'' through analytics and decision making. IoT environments can generate and consume vast amounts of data. But managing this data effectively and gaining meaningful insights from it requires us to understand its characteristics. Traditional scientific, enterprise and big data management approaches may not be adequate, and have to evolve. Further, these characteristics and the physical deployment environments also impact the quality of the data for use.   
In this paper, we offer a taxonomy of IoT data characteristics, along with data quality considerations, that are constructed from the ground-up based on the diverse IoT domains and applications we review. We emphasize on the essential features, rather than a vast array of attributes. We also indicate factors that influence the data quality. Such a review is of value to IoT managers, data handlers and application composers in managing and making meaningful use of data, and for big data platform developers to offer meaningful solutions to address these considerations.
\end{abstract}


%

\section{Introduction}
The Internet of Things (IoT) is an emerging form of cyber-physical infrastructure that integrates sensing and actuation, communication, computation and analytics to impact diverse applications.  
Traditional means of observation and control within scientific and engineering disciplines use single large instruments or hundreds of bespoke connected devices for monitoring of natural and physical systems. IoT transforms and scales this by having thousands or millions of affordable sensing and actuation devices that are deployed widely to monitor and manage infrastructure (power grids, manufacturing equipment), physical spaces (urban pollution), or even personal spaces (smart watch, fitness device). These devices use pervasive commodity networks, such as the public Internet, 3/4G cellular or low power wireless networks like LoRa, for communication. 

This gives unparalleled access to environmental and infrastructure data that will soon dwarf the data collected by enterprise systems and online platforms. As consumer, industrial and utility devices become inter-connected, data serves as the key commodity that is generated by and consumed to provide the ``smarts'' to the sensors and actuators in the IoT ecosystem~\cite{Ramchurn:2012:PSS:2133806.2133825}. Also, IoT data are often born-digital. Cyber-Physical-Social systems where data from enterprise/government (cyber), domain infrastructure (physical) and humans/their online agents (social) will be fused and integrated. This forms the source for extracting insights and a means to communicate outcomes among distributed entities~\cite{karkouch:2016}.

The diversity of IoT data and its potential value makes it important to understand its characteristics. Hensley, et al.~\cite{hensley:2014} divide the life-cycle of an IoT data element into the following stages:
\begin{itemize}
	\item \emph{Creation}: Things/sensors generate the data in response to some physical variable of their environment.
	\item \emph{Curation}: The data is transmitted and assembled at other (usually more ``intelligent'') components, over the network.
	\item \emph{Transformation}: Process, transform and act upon the data.
	\item \emph{Archival}: If the data can be used further or is necessary to be retained, it is stored in some repository.
	\item \emph{Deletion}: If the data object is of no further use, it is deleted.
\end{itemize}
There are various aspects of IoT data that impact their role in the above life-cycle, and affect how we manage and use the data for actionable insights. Understanding of such behavior is essential for platform, application and analytics developers to design effective tools and platforms, and build meaningful models and decision support systems. Further, the usefulness of data to an IoT application is only as good as the quality of the data. However, the consequences of these data-driven decisions can be significant, e.g., affecting power grids or patient health. Quality does not exist in a vacuum, but is tightly coupled to the application's needs. Hence, examining the quality dimensions of IoT data, and their relevance to diverse applications is another gap we address here.

This paper presents a taxonomy of IoT data characteristics, based on a survey of IoT domains and literature, and identifies intrinsic and domain-specific features of IoT data. In particular, it highlights dimensions of IoT data quality and factors that can affect it. Given that the application's
decisions rely on these datasets, 
understanding the quality parameters helps design effective solutions. 

A key distinction we draw from other such reviews~\cite{qin:jnca:2016,karkouch:2016} is to not be exhaustive for its own sake but to be relevant to developers and practitioners of IoT. Many studies build upon prior works about characterizing data, and its quality at large~\cite{philipchen:2014,wang:1996}. As a result, they offer hundreds of attributes that are often just of academic interest. We take a ground-up approach by examining several exemplary IoT domains and applications, and drawing our features directly from them. We also draw upon our prior experiences with IoT domains~\cite{simmhan:cise,simmhan:spe:2018} and scientific data management~\cite{simmhan:record:2005,simmhan:quality}. This enhances the value of our review, and also helps to draw similarities with the challenges faced in eScience, enterprise and web domains.

Specifically, we make the following contributions:
\begin{enumerate}
  \item We review representative IoT applications from diverse domains to motivate the need for this review, in Sec.~\ref{sec:iot:apps}.
	\item We propose a generalized taxonomy for IoT data characteristics in Sec.~\ref{sec:iot:data} based on observations from these domains and literature.
	\item We identify the dimensions of data quality relevant to IoT data and the factors that affect it, in Secs.~\ref{sec:iot:qual} and~\ref{sec:iot-factors} respectively.
\end{enumerate}


\section{Related Work}
\label{sec:rel:works}
There have been several studies that have examined various aspects of IoT data, 
and some have also identified its quality characteristics. We next discuss several of these studies, and contrast them with our current review.

\textbf{IoT Data and its Quality.} Qui, et al.~\cite{qin:jnca:2016} propose a taxonomy for IoT data along with a discussion of research efforts in IoT, such as streaming techniques, data modeling, and event processing. The data taxonomy presented classifies IoT data characteristics according to their relevance to each category -- Data Generation, Data Interoperability and Data Quality. However, the paper does not clearly define each of these aspects, and offers rudimentary examples. It briefly mentions data quality concerns as an open challenge, without substantiating it with linkages to applications that determine and are affected by the quality. 
We offer a more detailed structure and dimensions that are based on various application case studies and existing literature. This serves as a more in-depth view of data quality issues from the application's perspective.

A survey on data quality in IoT~\cite{karkouch:2016} provides a comprehensive review of the quality metrics. It reviews the basic characteristics for IoT data, and provides a high-level classification of the quality dimensions -- Intrinsic, Contextual, Representational and Accessibility. Subsequently, it maps the quality taxonomy from Wang, et al.~\cite{wang:1996} to the IoT realm, with the prominent aspects being
accuracy, timeliness, completeness, volume, and confidence. 
But in the process, they fail to capture the generalizable semantics of data and quality characteristics from an IoT application's perspective, and this make it practically less useful.

Yet another article focuses on an IoT data architecture that is sensitive to security and quality~\cite{sicari:2016}. As part of this work, the authors highlight certain limitations in the widespread use of IoT data, with management of data quality being a prime concern. They offer accuracy, completeness, privacy, security and trustworthiness as key data quality dimensions that they manage as part of their system.
Similarly, IoT data characteristics are discussed in passing elsewhere~\cite{sicari:2016}, with three dimensions of quality identified -- Accessibility, Ease of manipulation and Representation. They also discuss the challenges that stem from the fact that data changes over time, and it might even change its structure.
Similarly, Cai, et al.~\cite{cai:jiot:2017} also provide a brief overview of IoT data characteristics. Unlike us, these works lack in depth as characterization of data or its quality is not their primary goal, but rather a side-effect of designing IoT data systems. 

\textbf{IoT as Big Data.} The sources of IoT data tend to be streaming in nature, generating data in real-time from sensors. IoT data is a form of fast-data, or high velocity data. As a result, studies that review big data and their quality dimensions can influence our characterization effort. The commonly used ``3 V's'' of big data identify volume, velocity and variety of data~\cite{3vs}.
Veracity and value are sometimes included as additional ``V's'' of the dimensions~\cite{veracity-value}. These have a bearing on our IoT data taxonomy, but we are not limited by these. Further, the veracity dimension is a form of data provenance~\cite{simmhan:record:2005} which we highlight among our factors influencing data quality. Others have expanded on the big data characteristics further as well~\cite{philipchen:2014}.

Quality of streaming data has been examined earlier~\cite{klein:2009}, 
which offers a contrast with static data and models the propagation of quality along the stream. It uses accuracy, confidence, timeliness, completeness and context in their model and offers a quantitative function over these. 
Merino, et al.~\cite{merino:2016} 
define adequacy (fitness for use) of data based on constituent quality dimensions, 
and use ISO 25024:2015 data standards for evaluating the quality. 
Our prior work~\cite{simmhan:quality} has used provenance as a means to estimate the data quality, using various metadata factors and expert-defined functions over the data and transforming processes. Provenance is used to recurse over the lineage and incrementally build a quality score.
Others~\cite{thota:2017} 
identify causes of poor data quality and suggests methods to improve effective data quality management from an enterprise's perspective. 


\section{IoT Applications and the Role of Data}
\label{sec:iot:apps}
The IoT paradigm impacts a wide array of domains. This ability to fundamentally transform the operational aspects of multiple domains, and also open up novel applications lies at the heart of the popularity of IoT. 
Here, we explore the role of data from IoT in four canonical and diverse application domains -- \emph{smart cities, smart healthcare, smart farming, } and \emph{smart industry}. We then use these applications to motivate the characteristics of IoT data and its quality in subsequent sections.

\subsection{Smart City}
As the global population migrates to urban centers, research and development to enhance the livability of large cities becomes essential~\cite{DBLP:conf/eScience/SinnottBGGKMMNPTSSVW12}. Smart city is a broad term that indicates the use of technology to improve the quality of life of urban residents, and to make the delivery of public services more efficient. 
It encompasses several sub-domains, driven both by the city, such as smart transport and smart utilities, and by residents through smart homes and buildings~\cite{neirotti:2014}. 
Exemplars of such smart city developments include Barcelona's Intelligent City (BCI) project~\cite{gea:2013}, Singapore Intelligent Nation, and Rio de Janeiro Smarter City (Brazil)~\cite{angelidou2017role}.

\subsubsection{Smart Utilities}
``Smart'' delivery of public utilities such as power, water and sanitation is designed to improve the efficiency, cost, sustainability and safety of these services~\cite{Ramchurn:2012:PSS:2133806.2133825,simmhan:spe:2018}. A \emph{smart power grid} is one of the largest instantiations of the IoT network as part of the physical infrastructure. It integrates the classical power grid with large scale Information and Communication Technologies (ICT) and renewable energy~\cite{bekara:2014}. This includes monitoring and control of generation, transmission and distribution networks, all the way to integrating with smart home appliances and building area networks~\cite{yan:2013}. 
%
This introduces heterogeneity in the variables monitored, sensor models, and the management domains. These sources can also be physically distributed across 100's of kilometers, with implications on the network communications. 

Such data sources are used for near real-time applications like anomaly or outage detection, power stability, demand-response optimization, and spot market trading in renewables~\cite{gungor:2010,simmhan:cise}. Decisions made by such applications using data collected from the IoT fabric have serious consequences such as grid failures. Such datasets may also need to be retained for years for regulatory compliance, or for building better prediction models~\cite{Aman:tkde:2015}. Security and privacy of data from smart grids is also a concern, given the rise in cyber-physical attacks by non-state actors to bring down the grid, or leaks of real-time power consumption data that allows remote monitoring of household activity patterns
~\cite{parikh:2010}.



\subsubsection{Smart Transport}
Smart or Intelligent Transport domain 
spans a variety of ubiquitous vehicular and commute activities -- contemporary systems such as Automatic Vehicle Location Systems (AVLS)~\cite{debnath:2014}, smart parking and traffic management~\cite{gea:2013}, and road surface monitoring~\cite{syed:2012}, to more forward-looking aspects such as Autonomous Vehicles~\cite{OFRA1} and using Vehicles as Compute Infrastructure~\cite{hou:2016}. 

Data collection from sensors deployed as part of a smart transportation network have to contend with a dynamic environment. 
Vehicles need to communicate with other vehicles and roadside units to exchange information, and network connectivity can be intermittent~\cite{OFRA1}. 
The value of such data can also be immense since such data streams can be used for prediction and response, e.g., to control traffic signaling based on crowd-sourced trajectory monitoring~\cite{DBLP:conf/eScience/WangSN16}, or warning the driver about a likely tire burst. Such applications are highly time-sensitive, on the order of seconds or even milliseconds. 
This becomes even more critical with autonomous vehicles where processing delays due to data availability or network connectivity can cause safety hazards.
On the other hand, say, using GPSs for live tracking of bus 
transit are less time critical, on the order of minutes, while monitoring the road surface for fixing potholes~\cite{syed:2012} can withstand delays of days.


\subsubsection{Smart Homes and Buildings}
\emph{Building Management Systems (BMS)} are common-place in modern buildings, and allow digital monitoring and automated control over various infrastructure. 
Such ``smart buildings'' 
are able to centrally access and manage systems for electricity and water, Heating Ventilation and Cooling (HVAC), fire and emergency management, and access control and surveillance. 
They use a variety of sensors and actuators to enable these, and communicate using a LAN or WLAN backbone~\cite{OFRA4,minoli:2017}. 
These sensors can monitor at fine temporal and spatial granularities, 
say sampling the setpoint temperature and air flow speed of each HVAC unit every minute, or the CO levels of safety systems every few seconds. This leads to a large volume of real-time data used for alerts and online controls~\cite{dimakis:2010}. This can also be variable, e.g. access control systems are more active during the day but their outliers at night are more important. 

At the level of individual homes, several \emph{smart consumer devices and appliances} are designed to improve the quality of living. Smart washing machines and refrigerators can work with smart power grids for efficient energy management, while electric vehicles can communicate with smart grids to plan charging or return power back-to-grid~\cite{komninos:2014}. Devices like Nest integrate with home alarm systems and air conditioning units to offer remote access through apps. Having access to such diverse IoT data and controls can help intelligent meta-agents interface with these smart devices and control the various parameters of the smart home.
That said, the privacy of the resident is even more of a concern with such pervasive monitoring
~\cite{xu:2016}.

Another category of smart home applications have to do with \emph{assisted living} for senior citizens, for safety monitoring and cognitive and sensory support
~\cite{demiris:2008}. 
Fall monitoring systems use wireless motion sensors to detect interactions with the environment. These, when passed through event analytics, 
can detect and alert if the person has fallen down and needs assistance~\cite{banerjee:2017}.
However, faulty sensors or intermittent communications can cause false alarms or missed falls, both of which can be troublesome. 
Others have also used telemetry from wearables to identify activities that the elderly are performing to offer suggestions. E.g., the \emph{Ambient Kitchen project}~\cite{Pham:2009:SRF:1694626.1694632} is able to detect food preparation activities using accelerometer data and offer cognitive support.

\subsection{Smart Healthcare}
The use of wearables also has broader implications for healthcare delivery. 
Smart Healthcare includes applications such as patient monitoring, programmable drug delivery~\cite{OFRA3}, rehabilitation systems~\cite{fan:2014}, medication management, and smart phone-based healthcare solutions~\cite{islam:2015}. 
Specialized wearables, such as the ones used in the \emph{MOVEeCloud project}, track patient activity levels so that their primary care facility has comprehensive data when the patients go for a checkup~\cite{hiden2013developing}. However, such wearables can be heterogeneous and suffer from battery and accuracy issues
~\cite{hassanalieragh:2015,islam:2015}. 
Also, assigning semantics to the plentiful data can be challenging. 
Standards like IEEE 11073(x73) for health informatics specify communication protocols through a Data Distribution Service (DDS), and can be used for patient monitoring and drug delivery~\cite{OFRA3}. Here, automated systems 
administer required doses, eliminating human errors, but such systems themselves have to be robust. 
Efforts like the XPRIZE for a portable \emph{tricorder} device that can rapidly measure health parameters like ECG, eye and ear, and vitals of babies will make IoT-based healthcare data collection and service delivery ubiquitous even in developing nations~\cite{ackerman2015race}.

%
Smart Healthcare is a domain that has a little room for errors. Further, the data has to be maintained for long durations as part of patient, device and hospital history. Given 
the number of stakeholders, privacy is a key concern~\cite{asghar:2017}. 

\subsection{Smart Farming}
Smart farming is the inclusion of information and communication technologies in machinery, equipment, and sensors used in agricultural systems~\cite{pivoto:2017}. It stems from the idea of \emph{precision agriculture} that involves the use of sensors at the farm to monitor the soil and plant state, weather, disease outbreak, etc. and uses that information to make decisions on planting cycles, and control automated precision delivery of water and nutrients~\cite{DBLP:conf/eScience/DriemeierLPSFMF14}. 
E.g., \emph{Phenonet}~\cite{jayaraman:2015} is an agricultural phenotyping field laboratory that helps to assess the crop yield performance based on real-time capture of farm conditions like irrigation, climate and soil profile. 
%

Smart farming also depends on diverse and external factors like weather forecasts, soil profile studies, land use data, agri-market economics, etc.~\cite{kamilaris:2017} to make short term and long term decisions. 
The equipments and sensors 
used can have variable precision, ambiguities and poor interoperability~\cite{pivoto:2017}.  
Poor telecommunication infrastructure in rural areas can impact data collection. 
Also, farmers in developing nations may not 
have the proper skills to use the smart systems, and automated traceability of information 
becomes important. 
Misuse of this data by competing farms, or seed and insurance companies is also a concern
~\cite{nandyala:2016}.



These concepts extend to \emph{livestock farming} as well. 
The \emph{Internet of Animal Health Things (IoAHT)}~\cite{smith:2015} focuses on precision livestock farming and increasing their yield using IoT technologies. Sensors monitor the animals and the data collected helps identify and maximize the yield patterns. Data from such a system has other stakeholders such as feed providers, veterinarians, nutritional and pharmaceutical companies as well.

\subsection{Smart Industry}
Smart Industry, or \emph{Industry 4.0}, is another popular IoT domain~\cite{haverkort:2017}. It not only encapsulates smart factories but also smart machines and equipments like aircraft turbines~\cite{DBLP:conf/eScience/ElSaidWHD16}. This has implications on other IoT domains like smart grid for energy management as well. 
The industrial environment can be harsh, and requires robust IoT technologies as it can affect productivity and worker safety~\cite{OFRA2}. Malfunctioning sensors can lead to inaccurate and missing data. 
Lack of maintenance and other resource constraints are often present, 
requiring resilient models of data communication, management and processing
~\cite{haverkort:2017}. 
E.g., \emph{Real time In-situ Subsurface Imaging (RISI) for Oil and Gas Exploration}~\cite{OFRA2} processes oil well data in the field instead of sending it to the cloud due to limited remote communication. 
Mesh networks between edge and fog devices 
are used for data sharing and distributed decision making. 
There is also significant geographical spread 
that causes 
poor connectivity. 
The sensors are further limited by their battery life and maintenance is difficult due to the rough environment. 

\section{Characteristics of IoT Data}
\label{sec:iot:data}
\begin{figure}[ht]
  \centering \includegraphics[width=0.6\textwidth]{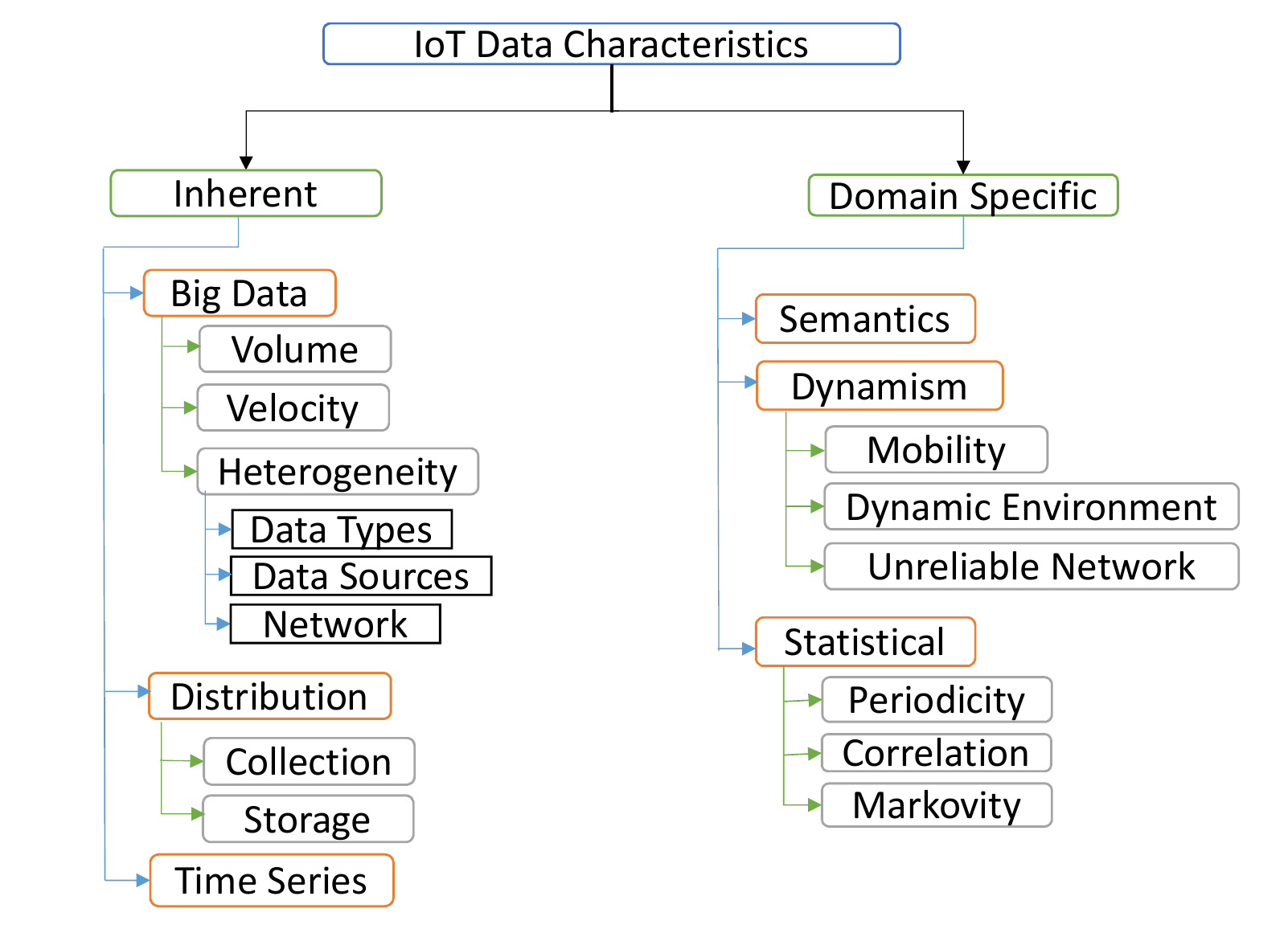}
  \caption{IoT Data Characteristics}
  \label{fig:iotdata}
\end{figure}

In this section, we propose a taxonomy for the characteristics of IoT data. 
These features are substantiated by one or more of the domains that we have discussed earlier.  
In particular, we emphasize on the characteristics of data that is \emph{generated directly from the physical IoT environment}, rather than focus on data that is post-processed by workflows or analytics at data centers. The suite of IoT dataflows and analytics is still emerging to offer a detailed treatment of the latter, and further, we expect them to share much of the characteristics of derived data seen in existing eScience and enterprise domains.

\subsection{Inherent Features}
We broadly categorize the IoT data characteristics into intrinsic features that we expect every IoT deployment to have, and domain-specific features that are present in specific application domains. 
We discuss the former here.

\subsubsection{Big Data}
IoT data is inherently a form of \emph{big data}. The widespread use of sensors for data collection, the use of the acquired data over long periods of time, and the need to analyze them at scale for decision support sees them overlap with several dimensions of big data~\cite{3vs}. 

\para{Velocity} The velocity dimension refers to how fast the data is being generated as well as how quickly the application needs to process it. ``Fast data'' is endemic to IoT 
since the data generated is a continuous and unbounded stream from a multitude of sensors, in most cases. Several factors affect the velocity of IoT data. It is directly influenced by the \emph{sampling rates} as well as the \emph{number of sensors} monitoring a particular environment~\cite{cai:jiot:2017}. E.g., phasor measurement units (PMU) in smart grids sample the power quality at $50-60$~Hz to ensure grid stability, and hundreds of these may be present in a network, with the need to respond in milliseconds. That said, there may be millions of consumer smart power meters that sample the demand every few minutes, or smart farming where such high sampling rates are rarely useful~\cite{jayaraman:2015}.

Sometimes, very high sampling rates might also lead to redundancy if the observations do not change rapidly. However, a certain amount of redundancy would be desirable for detecting the events with some level of confidence. In other cases, there may be a physical limit of the sampling rate due to the \emph{sensor sensitivity}. E.g., Particulate Matter (PM) sensors for air pollution monitoring measure the scattering of light pulses they emit. Their meaningful sampling interval is bound by the native pulse rate.

Also, since a primary goal of an IoT system is to detect changes in environment, 
it may be necessary to observe at higher rates to avoid missing transient events which might be outliers of interest.
E.g., applications like autonomous vehicles, due to their highly dynamic environment, produce data at a high velocity~\cite{OFRA1}. IBM estimates that a smart car continuously produces $1.3~GB$ of data every hour, and this may grow as we include higher definition video streams~\cite{fathy:2018}. Similarly, drug delivery systems~\cite{OFRA3}, fall monitoring systems~\cite{banerjee:2017} and emergency response systems~\cite{OFRA4} all generate data at a high sampling rate to detect anomalies and take decisions. Similar to sensors, one can also consider the \emph{control signals to actuators} and their related metadata as part of the IoT data stream~\cite{fathy:2018}. 
	
\para{Volume} This refers to the absolute quantity of data generated
~\cite{fathy:2018}. 
One view is to consider the unit volume of data generated per sensor and the number of (distributed) sensors~\cite{qin:jnca:2016}. 
But these sensors observe continuously, and this streaming data is potentially unbounded. It also depends on whether the data is being stored for archival, or just transiently retained and discarded after its use~\cite{cai:jiot:2017}. 

Velocity directly affects volume when data is accumulated and archived. E.g., in autonomous vehicles,  
the data volume collected for a day is $4~TB$~\cite{OFRA1}. But high velocity with low volume is possible if the data is not persisted, 
such as for a fall monitoring system where the high velocity accelerometer data 
can be discarded periodically~\cite{banerjee:2017}.
This also has an impact on the application performance and data availability
~\cite{simpson:2017}. When large volumes of high velocity data need to be pushed to a central location (e.g., cloud data center) for archival, this consumes network bandwidth and 
introduces unwanted latency~\cite{OFRA4}.

\para{Heterogeneity} The \emph{variety} aspect of big data indicates the diverse forms of data representations and protocols that are possible. Heterogeneity in IoT data is frequent, and can be due to different \emph{data sources}, \emph{data types} or \emph{network}.

IoT applications collect data from a wide variety of sensors, and in a wide variety of formats due to the diverse environment variables that are monitored 
~\cite{cai:jiot:2017}. Different data types (integer, character, images), representations (text, binary), formats (JSON, CBOR), models (structured, semi/un-structured), and vocabularies are possible~\cite{fathy:2018}.
E.g., a smart factory 
may have several types of sensors, 
each of which may use its 
own format to transmit the data. The form in which the data is stored may also be related or different from the source format as well. 
All of these can pose a challenge when the domain is inter-disciplinary, such as in smart grid~\cite{Zhou:itng:2012} or smart farming~\cite{jayaraman:2015}, where data from the sensors, weather, markets, social networks, etc. need to be integrated, and a common standard may not exist. 
In some domains, there may also be specific standards that are followed, such as medical devices using health informatics standards, but even these do not cut across other domains with which interaction is inevitably required.

Another view of heterogeneous data comes from the fact that the data collected can be both \emph{qualitative} (mostly contextual/semantics) and \emph{quantitative} (sensor values)~\cite{simpson:2017}. E.g., in a patient monitoring system, inputs from the doctors are collected as semantic annotations on top of objective data from sensors~\cite{banerjee:2017}.
%
Heterogeneity can be introduced by the sensors communicating using different protocols~\cite{fathy:2018} and network architectures. For example, certain constrained or legacy devices may not be able to support protocols like TCP for communications and instead use alternatives like UDP, SMS or modbus which may not be as reliable, or require specialized access and processing mechanisms
~\cite{abeele:2015}.

\subsubsection{Distributedness} IoT data and its sources tend to be spread spatially, depending on the region of interest~\cite{fathy:2018}. Applications like Smart Oil Fields~\cite{anand:2017} and Smart Cities~\cite{gea:2013} have sensor deployments spread over a large area. However, stand-alone smart building systems may have more densely packed sensors within small regions from where the data is \emph{collected}.
The geographical distribution of sensor placement impacts their communication over a network to where the data is \emph{stored and processed}. While this is often hosted at some centralized location like the cloud, edge and fog computing are emerging trends in IoT that allow for distributed storage and processing of the data as well~\cite{simmhan:iotn:2017}. These platforms are useful when the time and cost for data movement is high, such as for autonomous vehicles~\cite{OFRA1}, or privacy and regulations do not allow for their movement outside the data collection location, e.g., in smart healthcare.
When data collection and storage is distributed, processing may need to be moved where the data is present. Alternately, certain IoT applications may require access to multiple distributed data items, and in such cases a mechanism for data discovery, movement and integration is required. 

\subsubsection{Time Series} The IoT data is mostly observations from sensors, and hence tends to be time series data.
The IoT applications capture and use the information about the time at which data is generated. This brings up issues of synchronization of time across different (wide area distributed) devices, and differences between the time of observation and time of availability of the data. E.g., low end sensors may not have an on-board clock or may not have fine degrees of time synchronization using protocols like NTP. Some may use GPS or cellular networks to set their times. In certain cases, the time can only be assigned when the data is acquired, rather than when it is generated, and this can affect their use for the application consuming it.
%
IoT time series data can further be of two types, Static and Streaming~\cite{fathy:2018}.

\para{Static Data} Here, the data is accumulated and then transmitted in micro-batches or in batches for further processing. Hence, the total volume of the data remains constant while their availability is periodic. This may be due to the system design (e.g., the application does not require the data in real-time), to limit network transmissions (e.g., to conserve cost, bandwidth or energy), or because the communication network availability itself is periodic (e.g., using crowd-sourced ``data mules''~\cite{mishra:iotn:2015}).

\para{Streaming Data} Here, data is available for processing as it is generated. Data is transfered from the generation to the storage/processing location on the fly. 
Data here is potentially unbounded so it is either discarded after its use or is aggregated / converted in some compact form if storage is necessary (leads to creation of Static Data). 
This ``real-time'' may be near real-time, using best effort and accounting for network latencies, or truly real-time with strong transmission time guarantees. Mission-critical applications like 
patient monitoring~\cite{OFRA3}, autonomous vehicles~\cite{OFRA1}, and power transmission monitoring~\cite{gea:2013} require such hard limits. 

\subsection{Domain Dependent Features}
Here, we discuss IoT data characteristics that are relevant only to specific domains. 
\subsubsection{Semantics} Semantics provide contextual information about the data, and are metadata that add more meaning to the data collected. IoT sensors can generate large volumes of unstructured data without sufficient self-contained semantic information~\cite{cai:jiot:2017}. This is due to their constrained and commodity nature, and the context in which they are deployed. E.g., a simple temperature sensor can measure the ambient temperature of a living space or a gas turbine, and just report a binary observation value in both cases. But metadata such as the entity being observed, the units reported, the sampling interval, the error bounds, etc. can have serious implications when interpreting this value.
As is apparent, semantics are specific to individual domains. It is necessary for some, particularly if they operate across domains, but may be implicit agreed upon by the data sources and consumers in others.

Often, this semantic information has to be independently and explicitly collected and managed. Such information is necessary for IoT applications to make smarter and automated decisions
~\cite{qin:jnca:2016}. The type of semantic information captured can also span from RDF and OWL ontologies that allow reasoning~\cite{Zhou:itng:2012}, to a simpler glossary of well-defined vocabularies/units (e.g., SenML), to just unstructured text annotations (e.g., a PDF document of the sensor datasheet). 

\subsubsection{Dynamism} 
The data collection mechanism, period and even the values may vary from the usual set pattern due to different forms of dynamism 
-- \emph{Mobility}, \emph{Dynamic Environment} and \emph{Unreliable Network}.
Many IoT environments consist of several mobile components (sensors and/or compute devices), such GPS and other sensors in Electric Vehicles (EV) and public transport, or even autonomous vehicles in future. 
This adds dynamism to the data collection process as the spatial location changes the sensed parameters. Dynamism due to environment is a result of potentially hostile and unfavorable sensor conditions in applications like oil field deployments~\cite{anand:2017}, or vandalism of IoT devices at night in public spaces. 
The dynamism in data collection may even depend on the state of the device, such as an EV that is parked, one that is parked and charging, and the one that is in motion.
Lastly, dynamism due to 
poor and intermittent connections can cause data losses and/or delayed readings, especially when using wireless and ad hoc networks~\cite{qin:jnca:2016}. 
Some of these factors are related. E.g. mobility can cause network dynamism due to change in cellular signal strength. The sensitivity of applications to such dynamism and the ability for IoT platforms to mitigate or hide such variability is an important consideration.

\subsubsection{Statistical Properties} The data collected may have certain inherent statistical properties that can be exploited, say, for compression or validation.

\para{Periodicity} This is the simplest and most common statistical property that we can expect in many IoT applications like Smart Buildings that sample data at a fixed interval. 
Although the data collection is periodic, the data may or may not exhibit periodic patterns. E.g., diurnal and seasonal patterns in building energy usage helps build simple averaging and time-series forecasting models in smart grids~\cite{Aman:tkde:2015}, 
and data from ECG Monitoring System can be compressed for efficient storage~\cite{ukil:2015}.

\para{Correlation} 
IoT sensor data mostly have a timestamp and a sensor id, besides the observation, thus providing access to \emph{temporal} and \emph{spatial} domains to test spatio-temporal correlation~\cite{chen:2014:bigdata}. E.g., different copies of sensors in the same region may report similar values at a point in time. If multiple variables are monitored, there may be a correlation among them as well, such as the outside air temperature and the energy consumed by HVAC units.  
Such correlations can be used to draw insights and improve application performance~\cite{cai:jiot:2017}. 

\para{Markovity} Some physical quantities measured by the sensors are only a function of their immediate previous value, and do not depend on sensor value/state beyond that. 
Some notion of Markovity is always present in observational streams but establishing it is not as easy as just identifying patterns, probably because of the size of data~\cite{zachrisson:2017}. 
Such properties can help with better prediction models, or even be used to detect outliers that violate this property. 


\section{Dimensions of IoT Data Quality}
\label{sec:iot:qual}
\begin{figure}[t]
  \centering \includegraphics[width=0.8\textwidth]{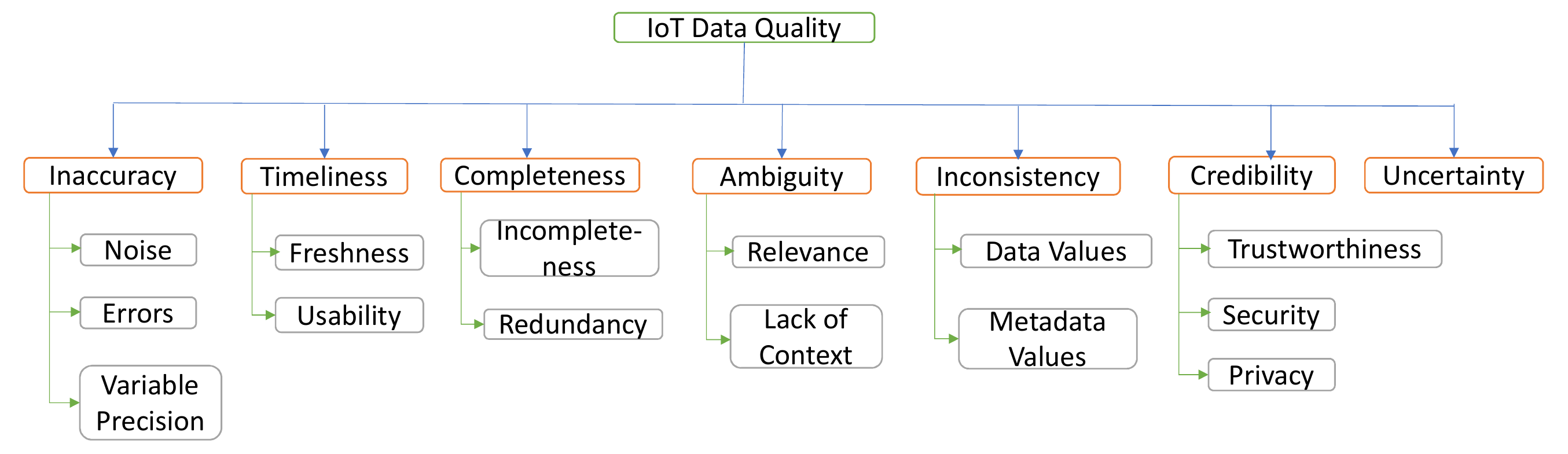}
  \caption{IoT Data Quality Characteristics
  }
  \label{fig:iotquality}
\end{figure}

Data quality is diversely defined as \emph{``the degree to which a set of inherent characteristics fulfills the requirements''}~\cite{banerjee:2017}, \emph{``how well data meets the requirements of user''}~\cite{karkouch:2016} and \emph{``the adequacy of data to the purposes of the analysis''}~\cite{merino:2016}. However, not all data quality attributes apply equally to IoT data~\cite{wang:1996}. To better understand quality and its relevance to IoT data, we have assembled a taxonomy of the most relevant quality dimensions here, motivated by the representative applications and IoT domains discussed in Sec.~\ref{sec:iot:apps}. These dimensions are illustrated in Fig.~\ref{fig:iotquality}.

\subsection{Inaccuracy}
Inaccuracy is the \emph{degree of incorrectness} of the data, and perhaps the most commonly used notion for quality that is widely mentioned.  
The importance of inaccuracy as a quality dimension for IoT is evident from the fact that most sensing systems can only capture 33\% of correct data~\cite{cai:jiot:2017}.

Inaccuracy has the following 3 aspects: 
\subsubsection{Noisy data}
Data values that do not follow the expected pattern contribute to noise. These can be erroneous values from faulty sensors or valid outliers. E.g., a sharp increase in room temperature of a smart building can indicate a fire emergency, while small variability can be noise due to doors opening and closing. The noise in data is important in applications where the sudden changes in the monitored variables are rare. In other words, the data possesses some statistical property like correlation or periodicity among data values or maybe even a Markovian pattern. Also, for data with high natural variability, we need to collect a large amount of data to discriminate noise from normal data trends, which means more data volume~\cite{simpson:2017}. 


\subsubsection{Errors}
Errors occur due to sensor failures, caused by age, harsh environments, or network data corruption.
In smart transport, the prime source of inaccuracy is the dynamic nature of data due to the mobile sensors~\cite{OFRA1}. 
Errors may also occur due to incorrect installation of the sensor, or its usage outside valid range of operations. 
E.g., industrial IoT requires more robust sensors than for home automation, though both may be observing similar variables.

\subsubsection{Variable Precision}
Sensors for measuring the same variable can have different precisions of observation, based on the sensitivity of their hardware and circuitry~\cite{chen:2014}. There can also be differences due to translation from analog to digital signals, where the specific algorithm used can introduce deviations. 
Some domains have standards that specify a required precision, but many may not. Precision is also related to accuracy -- a precise but inaccurate value is not useful. It is also possible that a large-scale IoT deployment, say in smart cities, can have similar sensors from different vendors but with slightly different characteristics. This heterogeneity can make it difficult to finely compare values across sensors.

\subsection{Timeliness}
The timeliness dimension is yet another key aspect of IoT data quality. Sicari, et al.~\cite{sicari:2016} refer to it as the age of the data that is acceptable for a task. IoT data is often generated continuously. Real-time or near real-time applications may require access to this in a streaming manner, and data that is buffered for too long may not be fit for processing.   
Thus, this dimension stems from the data being of high velocity and of a time series in nature. 
It can further be separated. 
\subsubsection{Freshness}
This is a quantitative measure of when the data is available for use, relative to it's generation. The freshness aspect is important to applications with low latency-tolerance, such as 
smart drug delivery and smart vehicles.  
E.g., when a vehicle is moving at a certain speed, data from forward facing sensors are useful only if available before a feature passes, i.e., forecast rather than nowcast, and this freshness threshold can vary based on the speed.

\subsubsection{Usability} 
This refers to how long a data can be retained before it becomes devoid of value. This assumes importance in a resource constrained device where processing all available data is both costly and unnecessary of the time to process exceeds the usable window of the data. Here, sampling techniques may be used. Dynamism also affects usability, say, if data is buffered due to network delays and too stale for use. 
Sometimes, a data's usability may be relative to whether newer data was already available from some other sensor. Usability is also applicable to making decisions on data archival, say, to train models in batch.

\subsection{Completeness}
The notion of completeness is a rather simple and intuitive one, but the definition varies from one use case to another. One study~\cite{sicari:2016} defines it as \emph{``the degree to which a given data collection includes data describing the corresponding set of real-world objects''}. 

\subsubsection{Incompleteness} 
A lack of completeness often indicates lower data quality. However, given that digital data is often a discretized view of an analog world, data from sensors is from sampling the environment and subsequent extrapolation. 
Thus, IoT data is inherently incomplete. But other forms of incompleteness can result from missing readings that occur due to lack of reliable communication, faulty sensors or occlusion of the observed entity~\cite{karkouch:2016}. 
This metric also requires that we know what completeness is, and can detect the lack thereof. If a sensor only reports values if there is a change, and incorrectly missed an observation indicating a change, then it may not be possible to detect it, leaving us with incomplete data.

\subsubsection{Redundancy}  
Redundancy refers to the presence of the same information more often than desired. This is important because it can help verify the correctness of data values and even rectify it. But it also results in wastage of network, storage, and processing. Redundancy can come from the \emph{spatial dimension} when sensors are geographically close to each other to be redundant, or \emph{temporal dimension}, when the sampling rate is much higher than the rate of change of the observations~\cite{qin:jnca:2016}.
IoT data can be incomplete even when it is redundant, e.g., when we have a lot of data about one particular event but missing data on some other event.

\subsection{Ambiguity}
The generated data can be interpreted in different ways, and different aspects of this data may be of use to multiple applications~\cite{qin:jnca:2016}. Ambiguity deals with how the data is interpreted by its consumer and application as well. 
\subsubsection{Relevance}
Relevance indicates how much of the received data is of use to an application. If the application receives data it has no use for or it was not expecting, it may result in ambiguity. 
E.g., in smart health care, providing the right data to doctors rather than all available medical data can help with better diagnosis~\cite{islam:2015}.
\subsubsection{Lack of Context} 
Sometimes, relevant information might still be ambiguous to an application because of a lack of context. This can be resolved to some extent by utilizing the semantic information about the data. But the contextual data is often unstructured in nature and hence more difficult to interpret automatically~\cite{simpson:2017}.
%
Another perspective comes from the mismatch between the provider and consumer of data. E.g., in smart farming,  
the field personnel may not have a complete understanding of the backend analytics 
and may misinterpret or ignore suggestions offered if it goes against their common wisdom~\cite{kamilaris:2017}.

\subsection{Inconsistency}
Inconsistency occurs when the same parameter measured at the same instant with different sensors results in different values. This indicates that either some of those values are not correct, or neither are, after accounting for precision and error bounds. Inconsistency can also cause ambiguity.
\subsubsection{Data values}
Different precision and sampling rates of the sensors can result in inconsistent readings. It can also result from network packet losses or sensor failures~\cite{qin:jnca:2016}. Applications in harsh environments with poor equipment maintenance tend to be more prone to such inconsistencies such as oil fields~\cite{OFRA2}. 
Inconsistency can also be used to detect and correct for failures and deviations if some sensors are known to be reliable reference sensors.
Sometimes, data corrections such as Kalman filters or interpolation that are applied to sensor outputs can also cause this mismatch. Inconsistent data should be treated as inaccurate data if not handled explicitly. 
\subsubsection{Metadata}
Inconsistencies can also occur in the data context or metadata. These result from improper annotations about the sensors or semantic mismatches. A direct impact of this is on the trustworthiness dimension as the metadata forms the sanity check by which credibility is determined.
Such concerns can arise when multiple entities maintain the same metadata, such as in smart healthcare where different caregivers may update a patient's record
~\cite{hassanalieragh:2015}.

\subsection{Credibility}

The concept of credibility of data can be further distilled into the following dimensions:
\subsubsection{Trustworthiness} This indicates how believable the data is~\cite{sicari:2016}. This often requires information on the trust in the data generator, subsequent operations performed on it, and even where and how it was stored. This makes it related to provenance as well, which is discussed later. 
Trust can be compromised either though omissions or commission. In the former, poor data curation practices can cause data to get unintentionally corrupted, such as in rigorous physical environments
~\cite{anand:2017}.  
There are also cases where trust is intentionally affected due to external agents, 
say, by modifying Smart Meters to under-report energy readings. 

\subsubsection{Security}
Security is about preventing unauthorized access to the data, and its sources, storage and processing units. Security is necessary to prevent intentional 
vandalism 
by unauthorized personnel that can also affect trust. E.g., 
healthcare and smart power grids are prone to attacks, and in future, autonomous vehicles are prime candidates. 
Having well defined security policies, means to enforce them, and ability to audit these will help improve credibility.

\subsubsection{Privacy}
Privacy ensures limits on who is allowed to access the data. While privacy may not directly affect credibility, having strong privacy guarantees enables data collection to be more transparent. E.g., patients may be willing to use wearables if they were confident that the data collected is kept confidential. 
Similar concerns exist in domains that intersect with humans and IoT technology, such as smart transport (spatio-temporal location), smart meters (resident activity patterns), etc. 
As it become easier to collect pervasive data, privacy is also necessary to prevent competitors from getting access to digital telemetry, say, on production output from smart industries.

\subsection{Uncertainty}
We define certainty as the probability or degree of confidence that we have on the available IoT data. This relates to the data being correct, complete, noise free, consistent and not subjected to alteration. Given the heterogeneity of IoT data and domains, uncertainty is sometime seen as inherent to IoT data~\cite{karkouch:2016}. In such cases, applications consuming the data need to take steps to address such uncertainty, either by including probabilistic processing and decision making, or by increasing the certainty through redundancies. 

\section{Factors influencing IoT Data}
\label{sec:iot-factors}
\begin{figure}[t]
	\centering \includegraphics[width=0.8\textwidth]{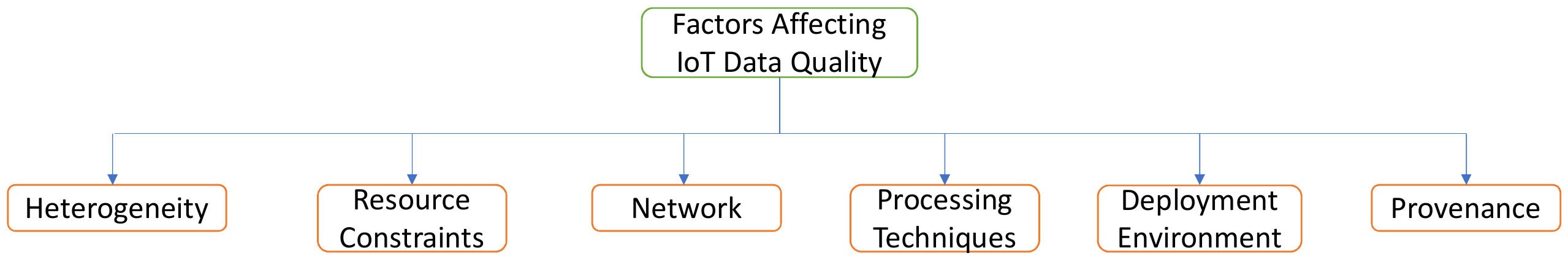}
	\caption{Factors affecting IoT Data Quality}
	\label{fig:factorsiotdata}
\end{figure}

As our final contribution, we discuss the factors that affect the quality of IoT data, and these can offer guidance on improving the quality or allowing the quality to be evaluated.
Unlike others~\cite{karkouch:2016}, we limit ourselves to  
the 
most relevant ones, as evidenced by the data characteristics, quality taxonomy and the domain applications. 
Fig.~\ref{fig:factorsiotdata} summarizes these factors. 

\subsection{Heterogeneity}
An IoT system is a complex construct of various infrastructures and devices and hence possesses significant heterogeneity in its ecosystem. As a result, data management and quality enforcement becomes difficult. 
This partly results from a lack of standards for IoT~\cite{weyrich:2016}, and thus makes data interoperability and a common understanding of quality difficult among different applications and domains.
Since IoT encompasses a very wide variety of domains and applications, it is very difficult to agree on a specific IoT design and deployment architecture~\cite{chen:2014}. Most architectures are tailored for specific applications~\cite{borgia:2014} and contribute to diversity. Interactions across IoT domains, say by verticals with cross-cutting horizontals, is also a factor that causes \emph{ambiguity} in the context and can challenge \emph{credibility}. 

Large IoT deployments also tend to have hardware from different vendors which hinders interoperability across various domains. 
These might not always be compatible and result in loss of data quality~\cite{miorandi:2012}. 
Such hardware heterogeneity can affect the \emph{accuracy} of the data, including variability in the precision. 
They can also lead to \emph{inconsistency} due to different models monitoring the same environment parameter differently. 



\subsection{Resource Constraints}
IoT sensors and micro-controllers often have limited resources -- energy, compute, memory and storage. 
This affects the data quality right at the acquisition stage, causing data loss (\emph{incompleteness}) due to limited buffers, bounded processing rates, and battery outages. Further, their commodity nature, necessary for affordable large-scale deployments, impact their sensitivity, leading to larger \emph{errors} and \emph{uncertainty}. 
Lastly, it may be difficult to configure these devices to run self-describing services or provide rich contextual metadata, instead relying on external sources for them. This can introduce a \emph{lack of context} in interpreting their data.


\subsection{Network}
``Things'' in an IoT deployment are connected through communication network, and these networks are integral to the IoT infrastructure. 
Diversity in the 
network protocols or network QoS can be caused by 
the variety of deployment models, ranging from mobile platforms to remote/rural environments, that preclude homogeneity~\cite{borgia:2014}. The number of networking standards themselves are vast.

The environments themselves may introduce dynamism 
in network coverage or performance, 
and cause applications to delay data transfers or drop packets -- affecting \emph{timeliness} or \emph{completeness}. 
Further, if constrained networks do not guarantee corruption free packets, we may also encounter \emph{inaccuracies}. 
As can be seen, the network design has significant bearing on many data quality dimensions.


\subsection{Processing Limits}
The constraints in the device hardware and the network also limit the type of software fabric that can run on the edge devices. 
It is possible that lossy preprocessing techniques like aggregation, sampling, etc. are employed to coarsen the data granularity~\cite{klein:2009b} and potentially degrade the \emph{precision} or \emph{completeness} of the data.
This reduction in data quality at the source can have a cascading effect on subsequent derived data, affecting their \emph{credibility}. 
Sometimes, there may be trade-offs between \emph{freshness} and \emph{completeness} -- where aggregate data is available in real-time but fine-grained data is available in daily batches.
Having high sampling rates may counter-intuitively cause quality to drop if all the redundant samples cannot be processed and some are dropped.

\subsection{Deployment Environment}
This factor captures the impact of the physical environment on the equipment used in IoT deployments.
Field deployments are often in rugged or hostile environments, exposed to ambient factors like heat and moisture that can affect the sensor and compute hardware, or unauthorized entities. 
The challenges may vary between smart city deployments, e.g., with vandalism by external agents; solar fields for smart renewables, e.g., with dust pollution and stray animals; or oil rigs with saline water and sparse networks~\cite{anand:2017}. 
This can affect the \emph{completeness} and \emph{correctness} of data, introduce \emph{inconsistencies}, and also affect 
\emph{credibility} if physical access by third-parties is possible. 

\subsection{Provenance}
\label{sec:role:prov}

Provenance is the derivation history or lineage of a piece of data, and the origin of it~\cite{simmhan:record:2005,davidson:sigmod:2008}. 
Provenance has been well studied in the context of eScience and collaboratory environments, particularly in the context of workflow-based processing.  
There has also been work on provenance for sensor and streaming data~\cite{hensley:2014,huq:DEXA:2011}.
In the past, provenance has been used as a means to ascertain the quality of scientific and linked data~\cite{simmhan:quality,Hartig:2009:UWD:2889875.2889881}. This has an impact on IoT data as well, given the distributed nature of its collection and processing, and the many institutions and domains involved.

The ability to gather provenance on IoT data can help determine and improve its quality, using existing or new techniques. Bauer, et al.~\cite{bauer:2013} even set forth requirements for provenance collection in IoT environments. 
Importantly, this can enhance the \emph{credibility} in the data, and the ability to trace across multiple sources, processing entities and their owners~\cite{cao:2016}. It can also offer \emph{context} to help resolve ambiguities and inconsistencies among the numerous sensors. Further, it helps with trouble-shooting where in the lineage chain quality degradations occurs~\cite{suhail:2016}.

However, gathering provenance data in a wide-area distributed and resource constrained environment is challenging, and can impact the performance of IoT systems~\cite{suhail:2016}.
The volume of (meta)data gathered may even outstrip 
the source data~\cite{bauer:2013}. Further, detailed provenance can also have \emph{privacy} implications, revealing knowledge about the data to unauthorized users. So this requires a careful trade-off between costs and benefits.

\section{Conclusions}
\label{sec:conclusion}
Internet of Things is arguably a vast source of born-digital data that can dwarf enterprise, scientific and web data. The use of such data is of societal value and mission critical in nature. It is also highly diverse. 
In this paper, we describe several representative IoT domains and applications, from the context of their generating and using data. The source is often the sensors and virtual devices, and the eventual consumers are analytics, digital agents, and actuators. We use these exemplar domains to identify intrinsic factors of IoT data as well as specialized features limited to certain domains. 
We then offer a similar taxonomy for characterizing the quality of this data and factors that affect it. This ranges from physical and environmental factors, to network and processing mechanisms. 
This contrasts from traditional enterprise data that rely on vertical integration, or web data that is publicly hosted and semi-structured, and has similarities with scientific data that is multi-disciplinary and often fed from instruments. This review can be of help to design better platforms and governance mechanisms to manage data and its quality.~{\let\thefootnote\relax\footnote{\textsc{Acknowledgment:} This work was supported in part by a research grant from NetApp, Inc.}}

\bibliographystyle{IEEEtran}
\bibliography{arxiv}

\end{document}